\documentclass[aps,prb,twocolumn,groupedaddress,floatfix,longbibliography,superscriptaddress]{revtex4-2}

\usepackage{amsmath,amssymb,graphicx,hyperref}

\usepackage{xcolor}
\usepackage{hyperref}
\usepackage{float}
\hypersetup{
colorlinks,
linkcolor={red!50!black},
citecolor={blue!80!black},
urlcolor={blue!80!black}
}

\renewcommand{\eqref}[1]{Eq.~(\ref{#1})} 
\renewcommand{\vec}[1]{\mathbf{#1}}
\newcommand{\ket}[1]{\left| #1 \right\rangle}
\newcommand{\bra}[1]{\left\langle #1 \right|}

\begin{document}


\title{Two qubit gate with macroscopic singlet-triplet qubits in synthetic spin-one chains in InAsP quantum dot nanowires}


\author{Hassan Allami}
\affiliation{Department of Physics, University of Ottawa, Ottawa, ON K1N 6N5, Canada}
\author{Daniel Miravet}
\affiliation{Department of Physics, University of Ottawa, Ottawa, ON K1N 6N5, Canada}
\author{Marek Korkusinski}
\affiliation{Department of Physics, University of Ottawa, Ottawa, ON K1N 6N5, Canada}
\affiliation{Security and Disruptive Technologies, National Research Council, Ottawa, Canada K1A0R6}

\author{Pawel Hawrylak}
\affiliation{Department of Physics, University of Ottawa, Ottawa, ON K1N 6N5, Canada}



\begin{abstract}
We present a theory of a two qubit gate with macroscopic singlet-triplet (ST) qubits in synthetic spin-one chains in InAsP quantum dot nanowires. The  macroscopic topologically protected singlet-triplet qubits are built with two spin-half Haldane quasiparticles. The Haldane quasiparticles are hosted by  synthetic spin-one chain realized in chains of InAsP quantum dots embedded in an InP nanowire, with four electrons each. The quantum dot nanowire is described by a Hubbard-Kanamori (HK) Hamiltonian derived from an interacting atomistic model. Using exact diagonalization and Matrix Product States (MPS) tools, we demonstrate that the low-energy behavior of the HK Hamiltonian is effectively captured by an antiferromagnetic spin-one chain Hamiltonian. Next we consider two macroscopic qubits and present a method for creating a tunable coupling between the two macroscopic qubits by inserting an intermediate control dot between the two chains. Finally, we propose and demonstrate two approaches for generating  highly accurate two-ST qubit gates : (1) by controlling the length of each qubit, and (2) by employing different background magnetic fields for the two qubits. 
\end{abstract}
\maketitle

\section{Introduction \label{sec:intro}}

There is currently interest in developing  robust qubits with long coherence times for quantum information processing on various platforms. Examples of  platforms include superconducting qubits based on macroscopic quantum states \cite{sc_qb_anneal_nature_2011,top_transmon_2011,sc_qb_nature_2017,google_quantum_sup}, trapped ions \cite{ion_trap_read_out_2008,ion_trap_wright2019,ion_trap_prx_2021}, quantum photonics \cite{loqc_2001,loqc_ph_qb_2007,xanadu_quantum_advantage,northup2014quantum}, and semiconductor spin qubits \cite{coded_qubit_chapter,nano_qi_2023,spin_qd_qubit_2013,coupled_qd_qb_1997,qc_qd_1997,so_qubit_si_2021,coherent_spin_qd_2006,store_spin_si_2014,Petta_STQB_2005}. The semiconductor spin qubits approach is particularly attractive for its promise of seamless integration with electronic devices. An important objective of any design is to achieve robustness against quantum noise. To that end, some strategies include realizing topologically protected qubits \cite{qd_sc_array_mzm_2012,mzm_review_2018}, using well-isolated qubits \cite{store_spin_si_2014,N_diamond_2014}, and encoding the qubit in composite structures \cite{coded_qubit_chapter,Petta_STQB_2005,spin_cluster_prb_2003,spin_bus_prl_2007,spin_bus_prb_2012,spin_bus_iop_2012,blazej_mdpi_2019}. A notable example of the latter is a qubit encoded in the singlet and triplet states of two electron spins in two gated lateral quantum dots \cite{Petta_STQB_2005}, a design that is protected against collective dephasing \cite{DFS_prl_97,DFS_QD_array}. Another related approach is to encode the qubit in two complex states of a spin cluster, thereby reducing the chance of a bit-flip error \cite{spin_cluster_prb_2003}.

Following these ideas, it was proposed to encode the qubit in the low-energy macroscopic quantum states of a synthetic antiferromagnetic spin-one chain \cite{Blazej_nature_2017,blazej_mdpi_2019,artificial_haldane_2010,artificial_haldane_2012,manalo_prb_2021,nanowire_qd_spec_2021,manalo_2024}. The low-energy spectrum of an antiferromagnetic spin-one chain consists of a singlet and a triplet separated by a gap from the rest of the spectrum \cite{HALDANE_pla_1983}, an example of topological phases of matter \cite{Haldane_nobel}. The topologically protected low-energy spectrum can be understood in terms of two spin-half Haldane quasiparticles localized at the two ends of the chain. A synthetic system of InAs quantum dots in an InP nanowire with four electrons each was proposed to realize such a chain \cite{Blazej_nature_2017,blazej_mdpi_2019,artificial_haldane_2010,artificial_haldane_2012,manalo_prb_2021,traingle_chain_2021,Catarina_Rossier,manalo_2024}, with the two spin-half quasiparticles resulting in a singlet-triplet Haldane (STH) qubit.

Previously, we have demonstrated how to realize the Haldane qubit in different physical systems \cite{artificial_haldane_2010,artificial_haldane_2012,Blazej_nature_2017,blazej_mdpi_2019,manalo_prb_2021,manalo_2024}. In this work, we demonstrate how to couple two STH qubits and generate two-qubit gates, opening the path toward universal quantum computation with STH qubits. To study various microscopic and effective spin Hamiltonians constructed throughout the paper, we used the exact diagonalization (ED), and MPS tools \cite{white_dmrg_prl_1992,mps_schollwock_2005} whenever the Hilbert space size was beyond the scope of ED.

The paper is organized as follows. We start in Section~\ref{sec:singlequbit} with a description of a single STH qubit realized in a chain of InAs quantum dots in an InP nanowire, for which we construct a Hubbard-Kanamori Hamiltonian and an effective spin-one Hamiltonian. In this section, we also briefly discuss how to generate all single-qubit gates using STH qubits. In Section~\ref{sec:two_coupled}, we construct a microscopic model for on-demand coupling of two STH qubits by an intermediate gated control quantum dot. Here we demonstrate that two coupled STH qubits, each made of two spin-half Haldane quasiparticles,  behave akin to two electronic ST qubits. Then in Section~\ref{sec:qubit_gates}, we discuss how to generate two-qubit gates using our proposed coupling scheme. Finally, in Section~\ref{sec:conclusion}, we summarize the results of our work and discuss future directions.

\section{Singlet-triplet qubit in synthetic spin-one chain \label{sec:singlequbit}}
\begin{figure}[ht!]
	\includegraphics[width=\columnwidth]{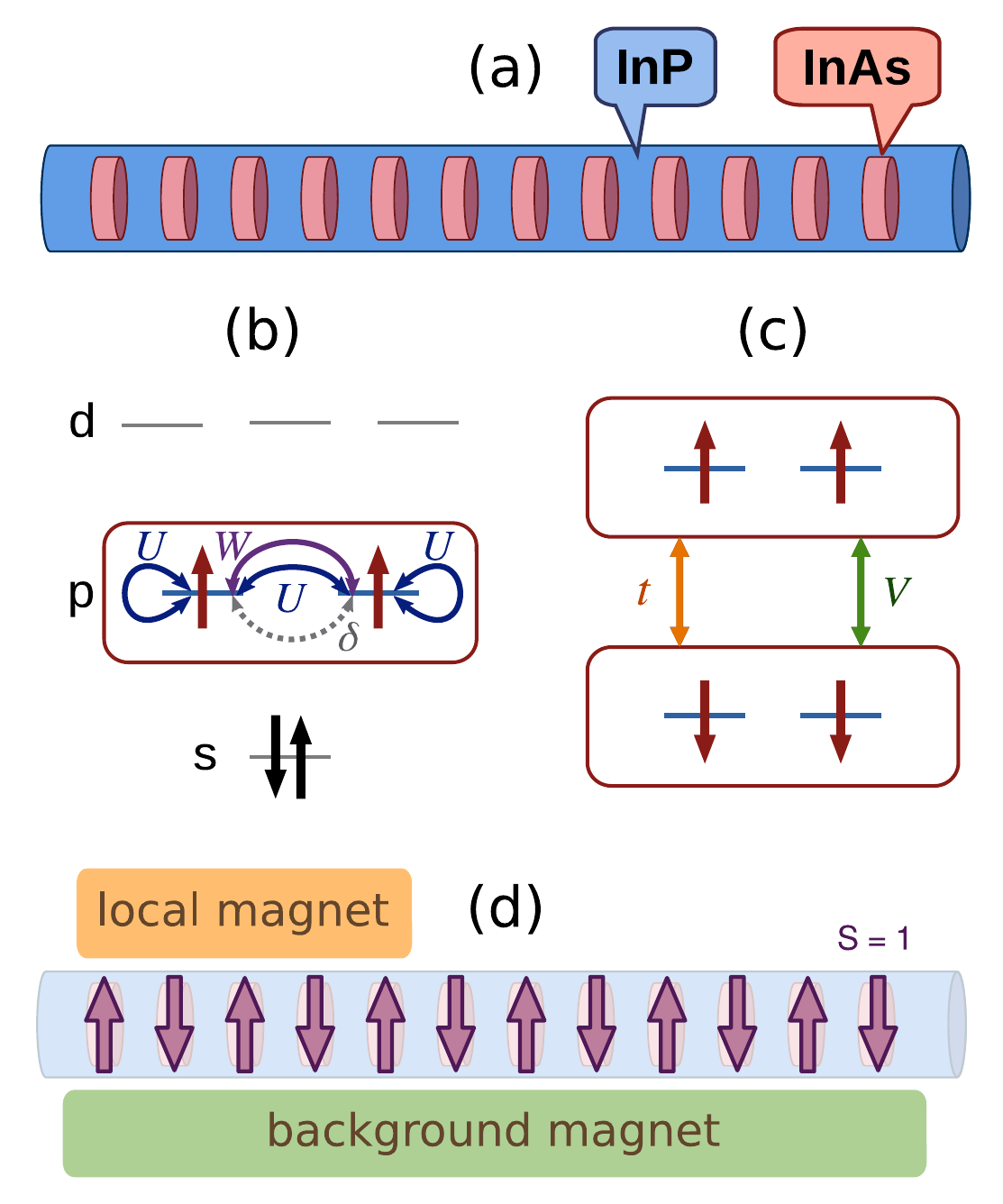}
	\caption{(a) Schematic of a synthetic spin-one chain as an array of InAs quantum dots in an InP nanowire. (b) The conduction band levels of each dot occupied by four electrons. Our model focuses on the two electrons within the half-filled p-shell, as described by \eqref{eq:H_intra}. (c) Schematic illustration of the p-shells of two adjacent dots as described by \eqref{eq:H_inter}. (d) In the low-energy limit, the system behaves like an antiferromagnetic chain of spin-ones, which can be utilized as a macroscopic singlet-triplet qubit. The background magnetic field moves undesired triplets with $S^z=\pm 1 $ away from the $S^z=0$ computational basis, and the controllable local magnet is used to generate single-qubit gates.}
	\label{fig:schematic}
\end{figure}

Let us start by describing the synthesis of an effective spin-one chain in an InAsP quantum dot nanowire system.
We consider an InP nanowire hosting a sequence of InAsP quantum dots. Experimental fabrication of such quantum dot nanowires has been successfully demonstrated, with theoretical investigations extending to atomic-scale details \cite{nanowire_qd_spec_2021,manalo_prb_2021}.
Fig.~\ref{fig:schematic}(a) shows a schematic view of such a quantum dot nanowire.

\subsection{Microscopic model of synthetic spin-one chain \label{sec:mirco_model}}

The atomistic microscopic model of InAsP quantum dot has been developed already \cite{manalo_prb_2021}. Despite hexagonal cross section and disorder due to low concentration of phosphor P atoms, the electronic states are grouped into shells similar to shells of a 2D harmonic oscillator, featuring $m+1$ orbitals corresponding to angular momentum $m$.
In Fig.~\ref{fig:schematic}(b), the conduction band shells of a dot are depicted schematically.
As the illustration shows, we populate each dot with four electrons, where two occupy the s-shell and the subsequent two occupy the p-shell.
We focus on these half-filled p-shell orbitals, which we call $p_\pm$, to construct a microscopic electronic Hamiltonian for the dot.
As we proceed to consider two such quantum dot nanowires, we designate the Hamiltonian describing dot $i$ in nanowire $A$ as $H_{A,i}$, given by
\begin{align}
	H_{A,i} &= U \left(
	(n_{A,i,+} - 1)(n_{A,i,-} - 1) + \sum_{\alpha=\pm}n_{A,i,\alpha,\uparrow}n_{A,i,\alpha,\downarrow}
	\right) \nonumber \\
	& - 2W \left(\vec{s}_{A,i,+} \cdot \vec{s}_{A,i,-} + \frac{1}{4}n_{A,i,+}n_{A,i,-}	\right) \nonumber \\
	& + \frac{\delta}{2} \sum_\sigma c^\dagger_{A,i,+,\sigma}c_{A,i,-,\sigma}
	+ c^\dagger_{A,i,-,\sigma}c_{A,i,+,\sigma},
	\label{eq:H_intra}
\end{align}
where $U$ denotes the Hubbard repulsion on each orbital, which has the same value as the direct Coulomb interaction between electrons on the two orbitals \cite{manalo_prb_2021}, $W$ is the Coulomb exchange between the electrons on two orbitals, and $\delta$ represents the splitting between the two orbitals induced by disorder and deviation from cylindrical symmetry.
All these terms are depicted graphically in Fig.\ref{fig:schematic}(b). Note that there is no onsite energy for the orbitals, indicating that the energy is measured from the $p_\pm$ levels. In writing \eqref{eq:H_intra}, we use 
$n_{A,i,\alpha,\sigma} = c_{A,i,\alpha,\sigma}^\dagger c_{A,i,\alpha,\sigma}$ and $n_{A,i,\alpha} = n_{A,i,\alpha,\uparrow} + n_{A,i,\alpha,\downarrow}$,
where $c_{A,i,\alpha,\sigma}$ denotes the annihilation operator of spin $\sigma$ for orbital $\alpha$ at site $i$ of the nanowire $A$. The spin operators are defined as
$s_{A,i,\alpha}^+ = c_{A,i,\alpha,\uparrow}^\dagger c_{A,i,\alpha,\downarrow}$, $s_{A,i,\alpha}^- = c_{A,i,\alpha,\downarrow}^\dagger c_{A,i,\alpha,\uparrow}$, and $s_{A,i,\alpha}^z = \frac12 (n_{A,i,\alpha,\uparrow} - n_{A,i,\alpha,\downarrow})$,
which then are used to build $\vec{s}_{A,i,+} \cdot \vec{s}_{A,i,-} = s_{A,i,+}^z s_{A,i,-}^z + \frac12 (s_{A,i,+}^+ s_{A,i,-}^- + s_{A,i,+}^- s_{A,i,-}^+)$.
The Hubbard repulsion $U$ prevents double occupation of the orbitals, while the exchange $W$ aligns the spin of the two electrons. Thus, unless the disorder $\delta$ is very large, the ground state of an isolated dot is a triplet. Consequently, in the low-energy limit, each dot can be deemed as effectively a spin-one object.

Next, we turn to the Hamiltonian describing the coupling between the p-shells of two adjacent dots in the nanowire $A$
\begin{align}
	H_{A, i,i+1} &= t\sum_{\alpha,\sigma}	\left(
	c_{A,i,\alpha,\sigma}^\dagger c_{A,i+1,\alpha,\sigma} + h.c. \right) \nonumber \\ 
	& + V(n_{A,i} - 2)(n_{A,i+1} - 2),
	\label{eq:H_inter}
\end{align}
where $t$ is hopping energy between the same orbitals and the same spins, and $V$ is direct Coulomb energy. Here we neglect the possible hopping between different orbitals and different spins, due to the symmetry of the wire, and the negligible spin-orbit interaction. And we only keep direct Coulomb matrix element as the other allowed Coulomb matrix elements are negligibly small \cite{manalo_prb_2021}.
In writing \eqref{eq:H_inter}, we also used the compact notation $n_{A,i} = n_{A,i,+} + n_{A,i,-}$. These terms are graphically shown between the two dots in Fig.~\ref{fig:schematic}(c).

Now, in the spirit of the Hubbard model, one would anticipate that the inter-dot hopping combined with intra-dot repulsion results in an effective antiferromagnetic coupling between the dots. In fact, second-order perturbation theory shows that for weak inter-dot coupling, the low-energy behavior of two dots is the same as two spin-one objects, coupled antiferromagnetically by an exchange energy given by $J_{\rm eff}=2t^2/(U+W-V)$ \cite{blazej_thesis}.

Combining the intra-dot \eqref{eq:H_intra}, and inter-dot Hamiltonians \eqref{eq:H_inter}, for a chain, we obtain a Hubbard-Kanamori (HK) electronic Hamiltonian that describes a quantum dot nanowire system of length $N$ as
\begin{equation}
H_A = \sum_{i=1}^N H_{A,i} + \sum_{i=1}^{N-1} H_{A,i,i+1}.
\label{eq:HK_A}
\end{equation}
Following the above discussion and results of exact diagonalization of a two quantum dot system with N=8 electrons, we find that in the limit of small inter-dot coupling, the HK Hamiltonian behaves like an antiferromagnetic spin-one chain \cite{manalo_2024}.

\subsection{Spin-one chain: \\ a topological singlet-triplet qubit \label{sec:single_spin_model}}

Now, we proceed to quantitatively confirm that in the limit of weak inter-dot coupling, the HK Hamiltonian behaves like an antiferromagnetic spin-one chain described by a Heisenberg Hamiltonian

\begin{equation}
    \widetilde{H}_A = J_{\rm eff} \sum_{i=1}^{N-1} \vec{S}_{A,i}\cdot \vec{S}_{A,i+1}.
    \label{eq:H_Heis_A}
\end{equation}

\begin{figure}[ht!]
	\includegraphics[width=\columnwidth]{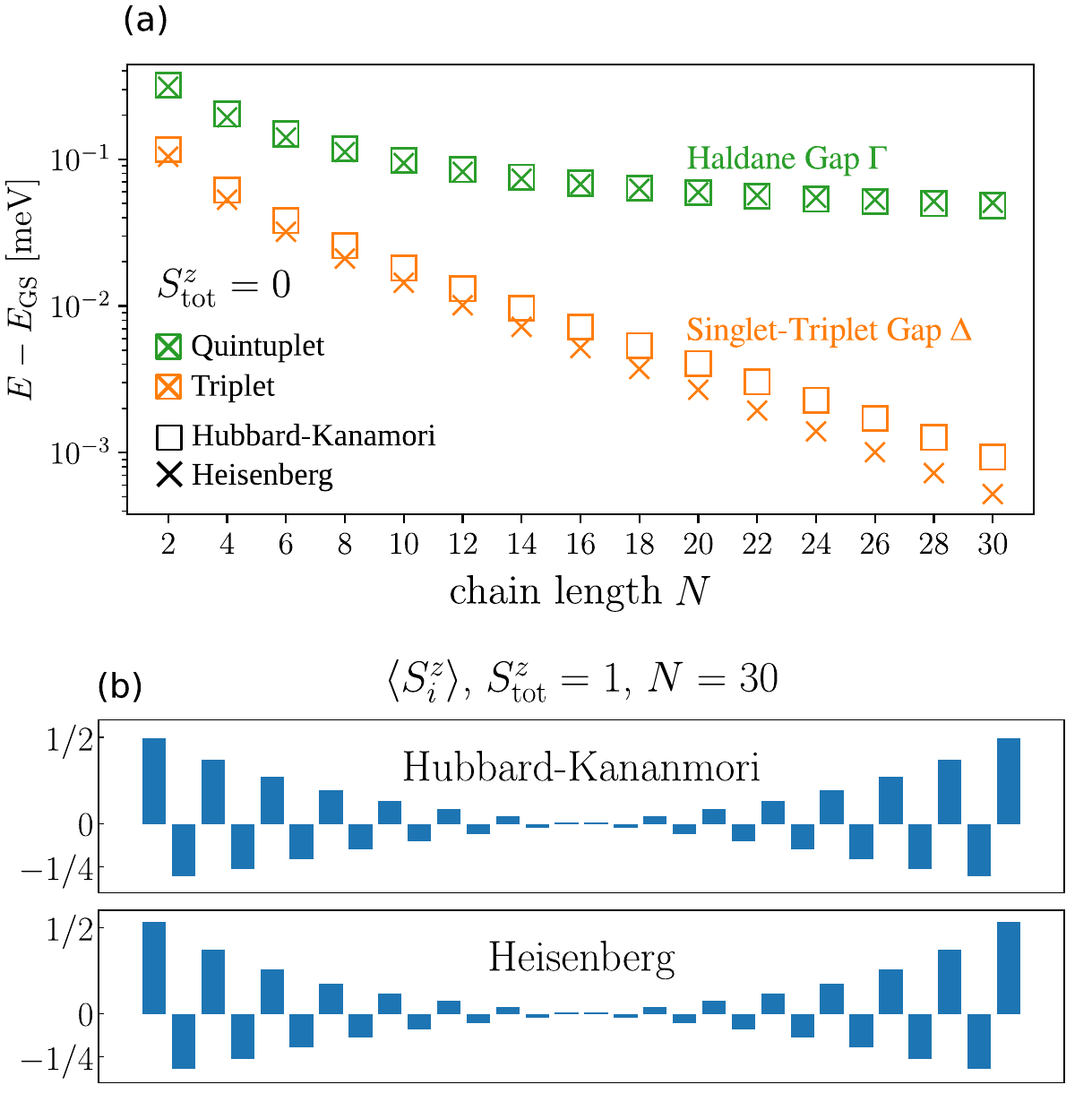}
	\caption{(a) The low-energy spectra of HK Hamiltonian of \eqref{eq:HK_A} with the parameters of Table~\ref{tab:HK_params}, and an effective spin-one Heisenberg Hamiltonian with $J_{\rm eff} = 2t^2/(U+W-V)\approx 0.1\text{ meV}$.
 (b) The expectation value of $S_i^z$ operator at each site $i$, over the lowest triplet state $\ket{T_+}$ with $S_{\rm tot}^z=1$, showing the spin-half Haldane quasiparticles, both for HK and Heisenberg chains of length $N=30$.}
	\label{fig:gap_sz_avg}
\end{figure}

For this demonstration, we employ the HK parameters listed in Table~\ref{tab:HK_params}, derived from an atomistic study \cite{manalo_prb_2021}.
To compare the low-energy spectra of the HK and Heisenberg Hamiltonians, we pick chains of even length, whose ground state is always a singlet.
In Figure~\ref{fig:gap_sz_avg} we present the results of our DMRG computation.
Fig.~\ref{fig:gap_sz_avg}(a) presents the low-energy spectra of the HK and Heisenberg Hamiltonians, measured from the ground state, for increasing chain length $N$. As anticipated for an antiferromagnetic spin-one chain, the low-energy spectrum includes a singlet, a triplet, and a quintuplet.
Haldane showed \cite{HALDANE_pla_1983} that this system possesses a topological gap. In particular, in the thermodynamic limit, the spin-one antiferromagnetic chain features a four-fold degenerate ground state comprising a singlet and three triplets, separated by a topological gap from the rest of the spectrum.
Fig.~\ref{fig:gap_sz_avg}(a) shows the exponential drop of the singlet-triplet gap $\Delta$ with chain length, while the topological Haldane gap $\Gamma$ converges to a constant value.
The spectrum of the HK closely mirrors that of the Heisenberg Hamiltonian, providing quantitative evidence that there exists a set of parameters for which the HK Hamiltonian behaves akin to a chain of spin-ones coupled by an effective antiferromagnetic exchange $J_{\text{eff}}$.

\begin{table}[ht!]
	\caption{ Parameters of the Hubbard-Kanamori (HK) Hamiltonian in \eqref{eq:H_intra} and \eqref{eq:H_inter}, based on a previous atomistic study \cite{manalo_prb_2021}.}
	\label{tab:HK_params}
	\begin{ruledtabular}
		\begin{tabular}{lcc}
            & Parameters & Values in meV\\
            \hline
            In& $U$ & 16\\
            Each& $W$ & 2.5\\
            Dot& $\delta$ & 0.85 \\
            \hline
            Between& $t$ & 0.75\\
            Dots& $V$ & 7.7\\
		\end{tabular}
	\end{ruledtabular}
\end{table}
The gapped spectrum suggests that one can utilize the isolated singlet and triplet states of a Haldane chain to construct a robust macroscopic singlet-triplet qubit, provided that the operation temperature remains below the Haldane gap \cite{Blazej_nature_2017}.
We refer to such a qubit as a singlet-triplet Haldane (STH) qubit. Similar to a regular ST qubit \cite{Petta_STQB_2005}, one can envision an STH qubit as comprising two spin-half quasiparticles.
The two spin-half quasiparticles are the emerging fractional particles of the Haldane phase on the two ends of the chain, where the topological phase has an interface with the trivial phase outside the chain \cite{aklt}.
As a way of visualizing these spin-half quasiparticles for a chain of length $N=30$, in Fig.~\ref{fig:gap_sz_avg}(b) we show the expectation value of $S_i^z$, the z-component of spin operator on site $i$, over the state $\ket{T_+}$, which is the lowest-energy state with $S_{\rm tot}^z=1$.
The spin-half objects are evident at the two ends of the chain for both the HK and Heisenberg Hamiltonian cases.
The nearly identical behavior of $\langle S_i^z \rangle$ in both cases provides further evidence of how closely the HK Hamiltonian mimics a spin-one Heisenberg chain in the low-energy regime.

\subsection{Generating single-qubit gates \label{sec:single_qubit}}
Before moving on to discuss how we couple two STH qubits to generate two-qubit gates, let us briefly describe how one can generate single-qubit gates using them.
Previously, it was shown that single-qubit operations on a STH qubit can be achieved by applying a local magnetic field on the first site of the chain \cite{Blazej_nature_2017}.
Here, we demonstrate that any local field that does not cover the entire chain is suitable for performing single-qubit operations, thereby relaxing the need for a spatially highly resolved and controllable magnetic field.

Consider the Heisenberg model of the qubit A, now with two sets of magnetic field as depicted in Fig.~\ref{fig:schematic}(d)
\begin{equation}
    \widetilde{H}_A = J_{\rm eff} \sum_{i=1}^{N-1} \vec{S}_{A,i}\cdot \vec{S}_{A, i+1} +
    b_A \sum_{i=1}^{N_b} S_{A,i}^z +
    B_A S_{\rm A, tot}^z,
    \label{eq:Heisenberg_A}
\end{equation}
where $b_A$ is a dynamic local magnetic field covering the first $N_b$ sites, and $B_A$ is a uniform background magnetic field covering the entire chain.
The computational basis of the STH qubit, much like regular ST qubits, consists of the singlet state $\ket{S}$, which we take to be $\ket{0}$, and the triplet state $\ket{T_0}$ with $S_{\text{tot}}^z=0$, taken to be $\ket{1}$. These states are separated from the rest of the spectrum by the Haldane gap $\Gamma$, which is of the order of $0.4 J_{\text{eff}}$. Meanwhile, the gap between them, $\Delta$, diminishes exponentially with the chain length (see Fig.~\ref{fig:gap_sz_avg}(a)).
Turning on the local magnetic field $b_A$ breaks the conservation of total spin $\vec{S}_{\rm tot}^2$, causing a mixing between $\ket{S}$ and $\ket{T_0}$, allowing for the generation of single-qubit gates. The uniform background magnetic field $B_A$ serves to push the other two low-energy triplet states, $\ket{T\pm}$, away from the computational basis, thereby reducing the errors  stemming from spin-flipping noises.
As we discuss below, the background magnetic field plays a more substantial role in implementing two-qubit operations.
Note that since the local field operation still conserves $S_{\rm tot}^z$, it does not cause leakage to $\ket{T_\pm}$ even in the absence of the uniform background magnetic field.

Since the local field breaks $\vec{S}_{\rm tot}^2$ conservation, it also mixes the computational basis with every other state in the $S_{\rm tot}^z=0$ subspace. But as long as $b_A\ll \Gamma$, such leakages remain negligible, as all other states are separated from the computational basis by $\Gamma$.

\begin{figure}[ht!]
	\includegraphics[width=\columnwidth]{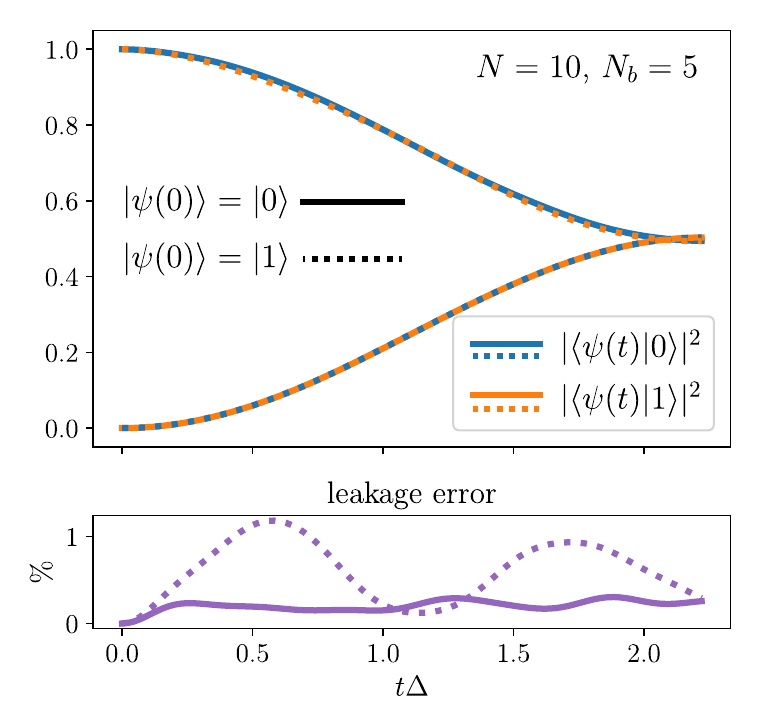}
	\caption{Implementation of the Hadamard gate using an STH qubit with a length of $N=10$, where the local magnetic field $b_A$ is applied to the first $N_b=5$ sites. The magnetic field is selected such that $M_b=\Delta/2$, where $M_b\approx 0.67 b_A$ is the matrix element of the local field between the singlet and the triplet. The solid(dashed) curves illustrate the qubit's evolution when initialized in state $\ket{0}(\ket{1})$. For each scenario, the time evolution of the square of the projections of the chain's state onto the computational basis $|\bra{\psi(t)}i=0,1\rangle|^2$ is plotted, along with (bottom) the leakage out of the computational basis, as indicated by \eqref{eq:leak_1}. The time on the x-axis is in units of $1/\Delta$.}
	\label{fig:hadamard}
\end{figure}

In Fig.~\ref{fig:hadamard} we demonstrate the generation of a Hadamard gate using a STH qubit of length $N=10$, where the local field is applied to the first $N_b=5$ sites.
For $N=10$ the singlet-triplet (ST) gap is $\Delta \approx 0.14 J_{\rm eff}$, the Haldane gap is $\Gamma\approx 0.76 J_{\rm eff}$, and the matrix element of the local magnetic field between the singlet and the triplet is $M_b\approx 0.67 b_A$.
Choosing $b_A$ such that $M_b = \Delta/2$, one can generate a Hadamard gate by applying the local field for $t_H = \pi/\sqrt{2}\Delta$ duration \cite{mike_ike_4}.

To generate Fig.~\ref{fig:hadamard}, we keep the first 100 states of $\widetilde{H}_A(b_A=0)$ to form a new basis for a reduced Hilbert space. Subsequently, we construct the Hamiltonian including the local field. The time evolution is then computed using $e^{-i t\widetilde{H}_A}$ in the reduced Hilbert space.
In the top panel, solid(dashed) curves show the evolution of the qubit initialized in the $\ket{0}(\ket{1})$ state. It is visible that in both cases, at the end of the operation, the system is in a nearly equal superposition of the $\ket{0}$ and $\ket{1}$ states, as expected for a Hadamard gate.
Note that when the qubit is initialized in $\ket{1}$, there exists a $\pi$ phase difference between the two projections of $\ket{\psi(t)}$, which is not  visible in the plot, as we plot the square of the projections for visual clarity.

As mentioned, there is a small probability of the system leaking out of the computational basis. The leakage is quantified by the projection of the chain's state $\ket{\psi(t)}$ outside the computational basis
\begin{equation}
\epsilon_1 = 1 - \sum_{i=0,1} |\bra{\psi(t)}i\rangle|^2.
\label{eq:leak_1}
\end{equation}
As the bottom panel of Fig.~\ref{fig:hadamard} shows, the leakage is well below $1\%$ when the qubit is initialized in the $\ket{0}$, and it reaches a maximum of approximately $1\%$ when initialized in the $\ket{1}$. This can be attributed to the smaller separation of the triplet state ($\ket{1}$) from the rest of the spectrum above the Haldane gap.

Notice that to implement the Hadamard gate we only needed to control the local magnetic field. Furthermore, with a fixed $\Delta$, we can also generate any phase gate by setting $b_A=0$. This implies that we can generate all the single-qubit gates required for universality by solely controlling the local magnetic field \cite{mike_ike_4}.

\section{Coupling two Singlet-Triplet Haldane Qubits \label{sec:two_coupled}}
In this section, we first describe how to make a tunable coupling between two STH qubits, A and B, each realized in a quantum dot nanowire system. Then, using the effective spin Hamiltonian of the two coupled STH qubits, we compare their coupling with that of two coupled simple electron based ST qubits, each consisting of two spin-half particles.

\subsection{Microscopic model of the inter-chain coupling \label{sec:coupling_micro_model}}
\begin{figure}[ht!]
	\includegraphics[width=\columnwidth]{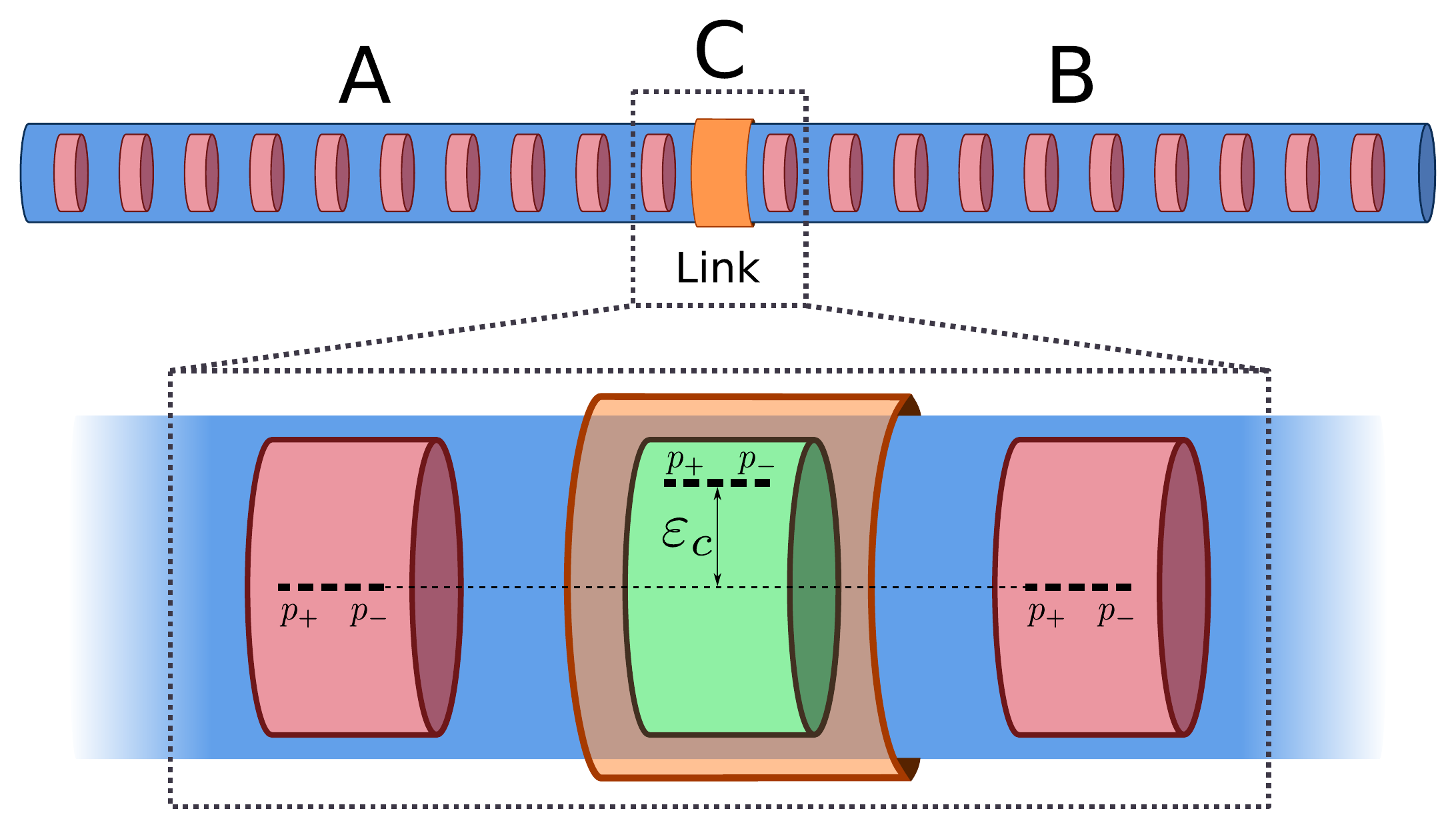}
	\caption{Schematic illustrating two STH qubits coupled via a link consisting of a gated control dot. The gate voltage $\varepsilon_c$ shifts the p-shells of the control dot relative to its neighbors.}
	\label{fig:the_link}
\end{figure}

Consider two STH qubits, A and B, coupled by a tunable link, as depicted in Figure~\ref{fig:the_link}. The tunable link is made of another quantum dot, labeled C and referred to as the ``control dot," which is gated by a gate tunable with applied voltage $\varepsilon_c$. As shown in Fig.~\ref{fig:the_link}, adjusting the gate voltage shifts the states of the control dot relative to its neighbors. When the p-shell of the control dot is out of resonance with their neighbors, the two ends of qubits A and B are effectively decoupled. Lowering $\varepsilon_c$, an effective coupling between the two ends of A and B develops, which, as we demonstrate, acts as a tunable effective antiferromagnetic spin coupling between the two ends of the two STH qubits.

The Hamiltonian describing the two coupled STH qubits is given by
\begin{equation}
    H = \varepsilon_c n_C + H_C +
    H_{AC} + H_{BC} + H_A + H_B,
    \label{eq:H_link}
\end{equation}
where $n_C$ is the electron number operator at the control dot, $H_C$ describes the control dot without the detuning $\varepsilon_c$ as in \eqref{eq:H_intra}, $H_{AC}$ and $H_{BC}$ describe the hopping and Coulomb interaction between the last dots of qubits A and B and the control dot as in \eqref{eq:H_inter}, and $H_A$ and $H_B$ are the HK Hamiltonians as in \eqref{eq:HK_A}, describing each qubit in isolation.

\begin{figure}[ht!]
	\includegraphics[width=\columnwidth]{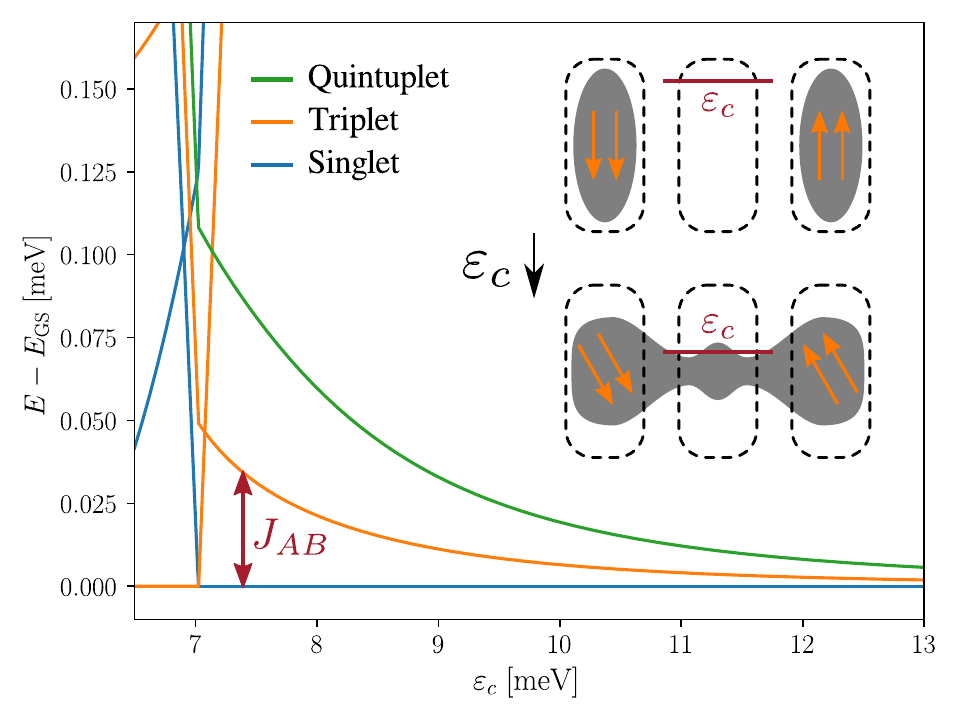}
	\caption{The low-energy spectrum of three dots, each populated by four electrons, as illustrated schematically in the inset. The gate voltage of the control dot, $\varepsilon_c$, varies along the x-axis. At high $\varepsilon_c$, the p-shells of the control dot are empty, and the two side dots are effectively decoupled. Decreasing $\varepsilon_c$ forms a bridge between the two side dots, antiferromagnetically coupling the two effective spin-ones on the side dots. The singlet-triplet gap gives the value of the effective coupling $J_{AB}$. Going below a critical value of $\varepsilon_c$, a phase transition occurs, and the system can no longer be viewed as two antiferromagnetically coupled spin-ones.}
	\label{fig:control_dot_coupling}
\end{figure}

To demonstrate that the link, realized by the gated control dot, functions as an effective tunable spin coupling between the two ends of the two STH qubits, let us focus on the link and consider three dots populated by four electrons. Two electrons belong to qubit A and two electrons belong to a dot of qubit B. Using the HK parameters listed in Table~\ref{tab:HK_params}, Figure~\ref{fig:control_dot_coupling} shows the low-energy spectrum of the three-dot system as a function of $\varepsilon_c$.
Recalling that two antiferromagnetically coupled spin-ones form a singlet, a triplet, and a quintuplet, one can observe that once $\varepsilon_c$ exceeds a critical value, the spectrum of the three-dot system resembles that of two antiferromagnetically coupled spin-ones. In this regime the changing singlet-triplet gap gives the value of the tunable effective spin coupling $J_{AB}$.
It is evident how this coupling quickly diminishes as $\varepsilon_c$ increases, allowing the two ends of qubits A and B to be perceived as two decoupled spin-one objects for large enough $\varepsilon_c$. Therefore, by adjusting $\varepsilon_c$, we can control $J_{AB}$ and effectively switch it on and off.

\subsection{Spin model of two coupled STH qubits \label{sec:coupling_spin_model}}

Let us now consider the effective spin Hamiltonian of two coupled STH qubits, A and B, given by
\begin{equation}
    \widetilde{H} = \widetilde{H}_A + \widetilde{H}_B +
    J_{AB}\vec{S}_{A,N}\cdot \vec{S}_{B,N},
    \label{eq:Heis_AB}
\end{equation}
where $\widetilde{H}_A$ and $\widetilde{H}_B$ are Heisenberg Hamiltonians, as described in \eqref{eq:H_Heis_A}, representing qubits A and B in isolation. The last term describes how the last dots of qubit A and B are coupled by the tunable coupling $J_{AB}$.

\begin{figure}[ht!]
	\includegraphics[width=\columnwidth]{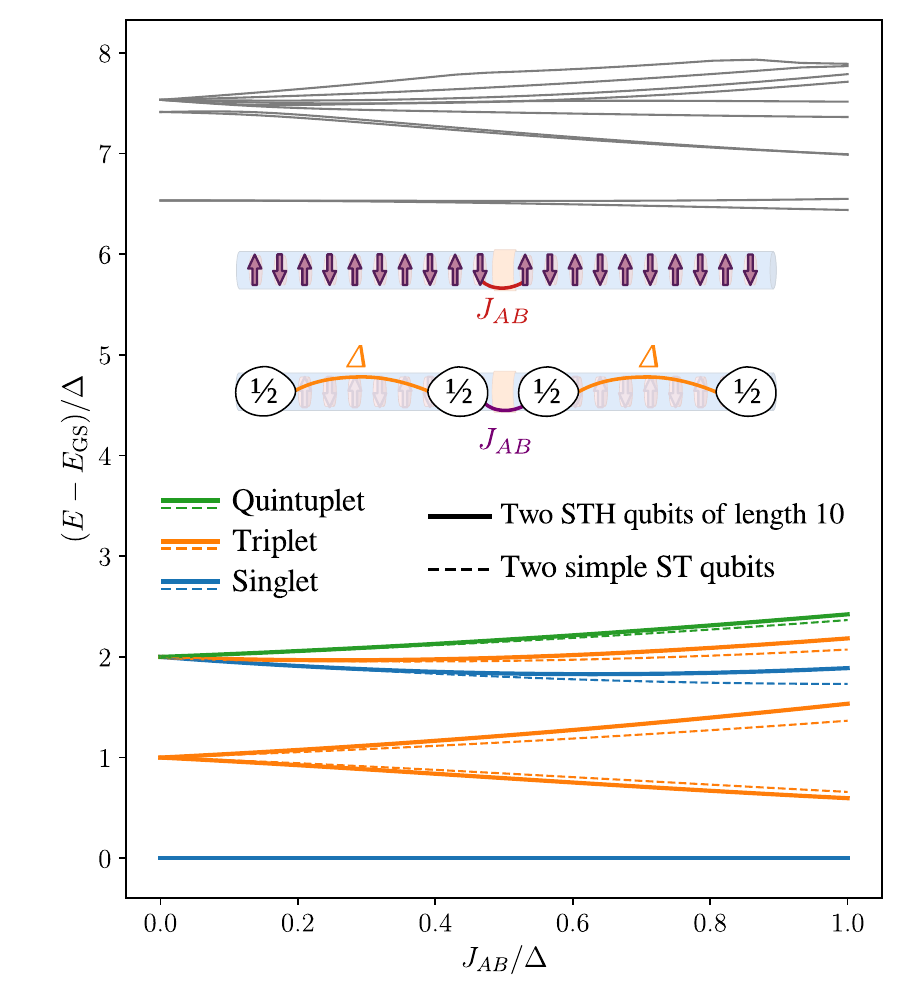}
	\caption{(solid curves) The spectrum of two coupled STH qubits of length $N=10$, described by $\widetilde{H}$ in \eqref{eq:Heis_AB}, plotted against $J_{AB}/\Delta$, where $\Delta\approx 0.14 J_{\rm eff}$ is the singlet-triplet gap of a single chain. Energies are measured from the ground state and are normalized to $\Delta$. The inset illustrates how STH qubits can be modeled as antiferromagnetic spin-one chains, with the low-energy spectrum of each spin-one chain viewed as two spin-half quasiparticles coupled by an exchange equal to $\Delta$. For comparison, the spectrum of two coupled simple ST qubits (dashed curves), described by \eqref{eq:HST_AB}, is also presented. The color of each curve indicates the total spin of the corresponding state. $\widetilde{H}$ exhibits higher-energy states beyond the six lowest energy states, separated from the low-energy manifold by a Haldane gap. The first few of these higher-energy states are shown as grey curves.}
	\label{fig:Heis_N10_coupled}
\end{figure}

As mentioned, the low-energy behavior of a single chain can be interpreted as two emerging spin-half quasiparticles coupled by an exchange coupling equal to the singlet-triplet gap, $\Delta$. In this perspective, which is schematically illustrated inside Fig.~\ref{fig:Heis_N10_coupled}, we anticipate that coupling qubits A and B by $\vec{S}_{A,N} \cdot \vec{S}_{B,N}$ will be analogous to coupling one spin-half quasiparticle of qubit A to a spin-half quasiparticle from qubit B.

On the other hand, two simple ST qubits, each composed of two spin-halfs, coupled in the same fashion can be described by
\begin{equation}
    \widehat{H} = \Delta (\vec{s}_{A,1}\cdot \vec{s}_{A,2} + \vec{s}_{B,1}\cdot \vec{s}_{B,2})
    + J_{AB}\vec{s}_{A,2}\cdot \vec{s}_{B,2},
    \label{eq:HST_AB}
\end{equation}
where $\vec{s}_{X,i}$ denotes the spin-half number $i$ of qubit $X$, $\Delta$ is the coupling in each ST qubit, and $J_{AB}$ couples the two ends of qubit A and B.
Since $\widehat{H}$ conserves the total spin, the computational basis of the two simple ST qubits evolves in the subspace $s_{\rm tot}^z=0$, which is a six-dimensional space spanned by $\lbrace \ket{SS}, \ket{ST_{0}}, \ket{T_{0}S}, \ket{T_{0}T_{0}}, \ket{T_{+}T_{-}}, \ket{T_{-}T_{+}}\rbrace$, where each component represents the tensor product of a state from qubit A and a state from qubit B.
Notice how the two components $\ket{T_\pm T_\mp}$ do not belong to the computational basis.
In this basis and with this ordering $\widehat{H}$ has the following structure
\begin{equation}
    \widehat{H}=
    \begin{bmatrix}
        H_{QQ} & H_{Q\overline{Q}}\\
        H_{\overline{Q}Q} & H_{\overline{QQ}}
    \end{bmatrix},
    \label{eq:HQQ_Tpm}
\end{equation}
with
\begin{subequations}
\begin{align}
    H_{QQ} &=
    \begin{bmatrix}
    -\Delta & 0 & 0 & \chi J_{AB} \\
    0 & 0 & \chi J_{AB} & 0 \\
    0 & \chi J_{AB} & 0 & 0 \\
    \chi J_{AB} & 0 & 0 & \Delta
    \end{bmatrix},
    \label{eq:H_QQ}\\
    H_{\overline{Q}Q} &=
    \begin{bmatrix}
    -\chi J_{AB} & \chi' J_{AB} & -\chi' J_{AB} & \chi'' J_{AB} \\
    -\chi J_{AB} & -\chi' J_{AB} & \chi' J_{AB} & \chi'' J_{AB}
    \end{bmatrix},
    \label{eq:H_qQ}\\
    H_{\overline{QQ}} &=
    \begin{bmatrix}
    \Delta -\chi'' J_{AB} & 0\\
    0 & \Delta -\chi'' J_{AB}
    \end{bmatrix},
    \label{eq:H_qq}
\end{align}\\
\label{eq:HST_blocks}
\end{subequations}
where the energy is measured from $\ket{ST_0}$ state, and for $\widehat{H}$ in \eqref{eq:HST_AB} we have $\chi=\chi'=\chi''=1/4$.
In Appendix~\ref{sec:coupled_ST_appendix}, we show how to analytically determine the spectrum of $\widehat{H}$  comprising two singlets, three triplets, and one quintuplet. This spectrum is plotted in Fig.~\ref{fig:Heis_N10_coupled} by dashed lines as a function of $J_{AB}/\Delta$.

Now, considering that the low-energy spectrum of $\widetilde{H}$ in \eqref{eq:Heis_AB} primarily consists of the low-energy singlets and triplets of the STH qubits A and B, and given that $\widetilde{H}$ also conserves total spin, its low-energy spectrum should be well-approximated in a similar six-dimensional basis, comprising the computational basis along with the two additional states $\ket{T_\pm T_\mp}$.
In Appendix~\ref{sec:matrix_element_appendix}, we provide a rigorous proof of why the projection of $\widetilde{H}$ onto this basis follows the structure shown in \eqref{eq:HST_blocks}. But for STH qubits, unlike the simple ST qubits, $\chi$, $\chi'$, and $\chi''$ are not identical, and they generally vary depending on the size of the chain. For two STH qubits of length $N=10$, they are approximately given by $\chi \approx 0.34$, $\chi' \approx 0.30$, and $\chi'' \approx 0.26$.

In Figure~\ref{fig:Heis_N10_coupled}, we present the low-energy spectrum of two coupled STH qubits with a length of $N=10$, described by $\widetilde{H}$ in \eqref{eq:Heis_AB}, computed using exact diagonalization. The solid curves represent the energy levels of $\widetilde{H}$ measured from the ground state as a function of $J_{AB}/\Delta$, where $\Delta$ is the singlet-triplet gap of a single chain, approximately given by $0.14 J_{\rm eff}$.
It is evident that the spectrum, normalized to $\Delta$, closely follows the spectrum of two simple ST electron spin based qubits described by \eqref{eq:HST_AB}.
The color of the curves corresponds to the total spin of the respective state, which is consistent for the lowest six states in both systems. The $S_{\rm tot}^z=0$ subspace of two coupled simple ST qubits is of course six-dimensional, but for two coupled STH qubits described by $\widetilde{H}$, the $S_{\rm tot}^z=0$ subspace is much larger, and contains many higher-energy states. However, these states are all separated from the six-dimensional low-energy spectrum by a Haldane gap. We display the first few of these higher-energy states as grey curves in Fig.~\ref{fig:Heis_N10_coupled}.

 \section{Generating two-qubit gates \label{sec:qubit_gates}}
Now that we have established how to couple two STH qubits, we proceed to demonstrate how to generate two-qubit gates using this coupling. It is well-known that for a quantum computer to be universal, it is sufficient to be able to generate all single-qubit gates, along with any two-qubit gate that is ``locally equivalent" to the CNOT gate \cite{mike_ike_4}. In Appendix~\ref{sec:CNOT_appendix}, we describe the concept of locally equivalent two-qubit gates and demonstrate how one can generate a gate locally equivalent to CNOT using a Hamiltonian with the structure of $H_{QQ}$ in \eqref{eq:H_QQ}.

\begin{figure}[ht!]
	\includegraphics[width=\columnwidth]{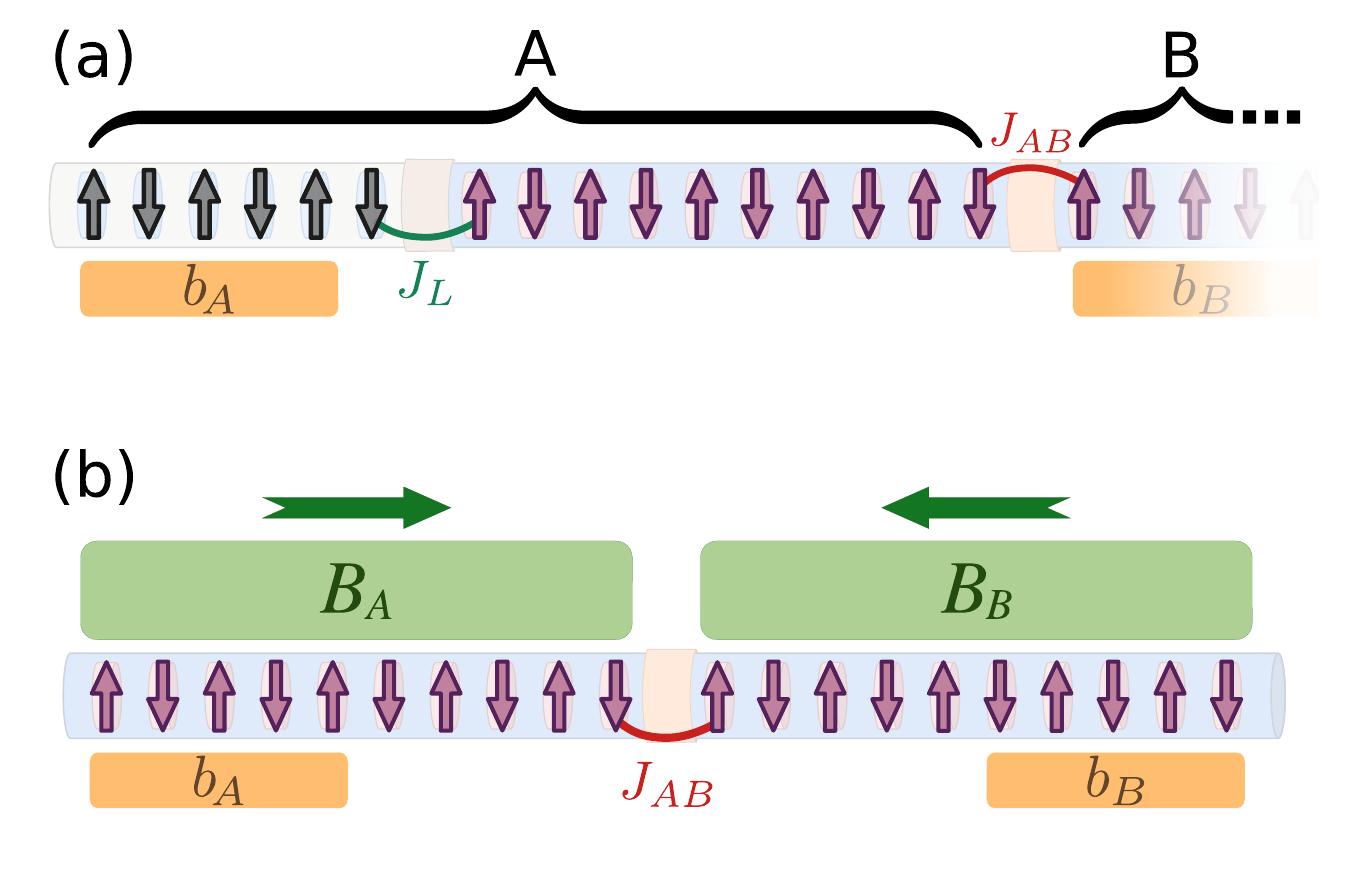}
	\caption{(a) Two STH qubits, A and B, with variable lengths coupled by $J_{AB}$. The local magnetic fields $b_A$ and $b_B$ are applied at different ends of each qubit, following Levy's prescription \cite{levy_two_qb}. By toggling $J_L$ on and off, we control the length of each STH qubit, effectively switching off and on $\Delta$, the singlet-triplet gap of each qubit. (b) Two coupled STH qubits with fixed lengths, hence fixed $\Delta$. The local fields required for generating single-qubit gates are applied at the same ends. By applying two different background magnetic fields $B_A$ and $B_B$ behind qubits A and B, the undesired states $\ket{T_\pm T_\mp}$ are pushed away from the computational basis, thereby reducing the leakage error.}
	\label{fig:two_qubit_gate_generation}
\end{figure}

The primary challenge in generating two-qubit gates from ST qubits is apparent from the previous section, where we presented the spectrum of two coupled qubits. This challenge arises from the two undesired states $\ket{T_\pm T_\mp}$, which couple to the computational basis. For conventional ST qubits, Levy addressed this challenge by devising a unique sequence to produce the controlled Z (CZ) gate \cite{levy_two_qb}, which is locally equivalent to the CNOT gate. In Levy's sequence, the undesired states fully decouple from the computational basis at the end of the sequence. Crucial to this sequence is (a) the application of one of the local magnetic fields on the coupled ends of the qubits and the other on the non-coupled one, and (b) the ability to decouple the spin-halfs comprising each ST qubit, that is to be able to set $\Delta=0$ in our language. The first requirement can be fulfilled for STH qubits as well, as illustrated in Figure~\ref{fig:two_qubit_gate_generation}(a). However, the second requirement presents a serious challenge, as $\Delta$ remains fixed for a given chain length in STH qubits.

To address this challenge, we can leverage the fact that the singlet-triplet gap $\Delta$ in Haldane chains decreases exponentially with the chain length. Therefore, by connecting and disconnecting a few more dots at the end of each STH qubit, we can effectively switch $\Delta$ on and off. This may be achievable by incorporating an additional controllable link within each STH qubit, labeled $J_L$ in Fig.~\ref{fig:two_qubit_gate_generation}(a). When $J_L$ is activated, the chain becomes longer, turning $\Delta$ off, and when it is deactivated, $\Delta$ is turned on. It is important to carefully design the internal link of each qubit so that when $J_L$ is activated, it accurately replicates the $J_{\rm eff}$ coupling between other dots of the chain, ensuring a properly uniform Haldane chain of increased length.

Demonstrating a Levy sequence on two STH qubits with variable lengths, sufficiently long to effectively turn off $\Delta$, poses a significant computational challenge, particularly in tracking the time evolution of a system with a massive Hilbert space. Additionally, designing and fabricating STH qubits with variable lengths could present its own set of challenges. Therefore, we propose a simpler scheme where $\Delta$ remains constant in each STH qubit. As we discuss below, this is achieved by introducing different background magnetic fields for the two coupled qubits.

\subsection{Using different background magnetic fields \label{sec:diff_magnetic_fields}}
Introducing two different background magnetic fields for the two STH qubits modifies the spin Hamiltonian describing the coupled qubits to
\begin{equation}
    \widetilde{H}_{AB} = \widetilde{H} + B_A S_{A, \rm tot}^z + B_B S_{B, \rm tot}^z,
    \label{eq:H_BAB}
\end{equation}
where $\widetilde{H}$, as given in \eqref{eq:Heis_AB}, describes the coupled qubits without background magnetic fields. The addition of the background magnetic fields further modifies the structure of $H_{\overline{QQ}}$ as
\begin{equation}
H_{\overline{QQ}} =
\begin{bmatrix}
    \Delta -\chi'' J_{AB} + B_{AB} & 0 \\
    0 & \Delta -\chi'' J_{AB} - B_{AB}
\end{bmatrix},
\label{eq:H_qq_BAB}
\end{equation}
with $B_{AB} = B_A - B_B$.
This implies that when the two background magnetic fields are different and $B_{AB}\neq 0$, the two undesired states $\ket{T_\pm T_\mp}$ are displaced from the $\ket{T_0T_0}$ component of the computational basis by $B_{AB}$.
The greater the $B_{AB}$, the lower the leakage of the computational basis into $\ket{T_\pm T_\mp}$.
However, it is important to be cautious about the higher-energy states, as they can come down due to the non-zero $B_{AB}$ and potentially interfere with the computational basis, resulting in leakage.
The first of such states is $\ket{Q_-T_+}$, comprising the quintuplet in the $S_{A,\rm tot}^z=-1$ subspace and $\ket{T_+}$ of qubit B. In the absence of coupling, $\ket{Q_-T_+}$ is positioned $\Gamma - B_{AB}$ above $\ket{T_0T_0}$, where $\Gamma$ is the Haldane gap, while $\ket{T_+T_-}$ is $B_{AB}$ above $\ket{T_0T_0}$, and $\ket{T_-T_+}$ is $B_{AB} - 2\Delta$ below $\ket{SS}$. Therefore, minimal interference from the outside of the computational basis can be achieved around the point where $\Gamma-B_{AB}= B_{AB} - 2\Delta \to B_{AB} = \Delta + \frac{\Gamma}{2}$.

\begin{figure}[ht!]
	\includegraphics[width=\columnwidth]{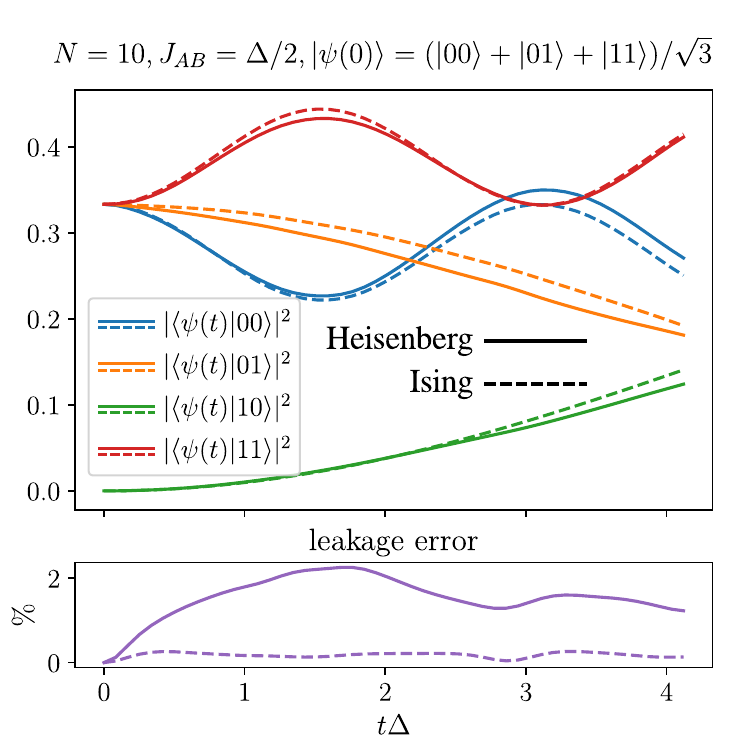}
	\caption{Time evolution of two coupled STH qubits each with a length of $N=10$, as governed by $\widetilde{H}_{AB}$ in \eqref{eq:H_BAB} (solid curves), with a coupling energy of $J_{AB}=\Delta/2$, half of the singlet-triplet gap of each qubit. Here, $B_{AB} = \Delta + \frac{\Gamma}{2}$, where $\Gamma$ is the Haldane gap of each qubit, selected to minimize the leakage error depicted in the lower panel, calculated using \eqref{eq:leak_2}. The two-qubit system was initialized in the state $\ket{\psi(0)} = (\ket{00} + \ket{01} + \ket{11})/\sqrt{3}$, and as $\ket{\psi(t)}$ evolves, the squares of its projections onto the four components of the computational basis are plotted in different colors. The duration of the evolution was chosen to be $t_c$, the time needed for a coupling pulse in the process of generating a gate locally equivalent to CNOT. This value was determined using the procedure outlined in Appendix~\ref{sec:any_J_AB}. Time on the x-axis is measured in units of $1/\Delta$. The dashed curves show the evolution and leakage of the same two qubits coupled by an Ising interaction.}
	\label{fig:Ising_Heis}
\end{figure}

In Figure~\ref{fig:Ising_Heis}, the solid curves show the evolution of two STH qubits, each with a length of $N=10$, coupled by $J_{AB}=\Delta/2$, and subjected to $B_{AB} = \Delta + \frac{\Gamma}{2}$. To generate this plot, we use the low-energy product states as the basis of the reduced Hilbert space, formed by the products of states from $S_{A,\rm tot}^z=m$ and $S_{B,\rm tot}^z=-m$. We progressively incorporate higher-energy states until the convergence of leakage, quantified by
\begin{equation}
    \epsilon_2 = 1-\sum_{i,j=0,1} |\bra{\psi(t)}ij\rangle|^2,
    \label{eq:leak_2}
\end{equation}
where $\ket{\psi(t)}$ is the state of the coupled system at time $t$, and $i$ and $j$ go over the computational basis.
In this case, we included 338 product states in the basis.

As the leakage panel of Fig.~\ref{fig:Ising_Heis} shows, the maximum leakage is about $2\%$. But note that one can always reduce the leakage by choosing smaller $J_{AB}$.
In Appendix~\ref{sec:any_J_AB}, we demonstrate that for any given $J_{AB}$, it is possible to create a gate locally equivalent to CNOT by applying the coupling $J_{AB}$ twice, with a local transformation in between. We also show how to find the duration of the coupling pulses, denoted as $t_c$, for any $J_{AB}$. In Fig.~\ref{fig:Ising_Heis}, we use the $t_c$ value obtained for $J_{AB}=\Delta/2$ as the duration of the time evolution.

To produce the time evolution depicted in Fig.~\ref{fig:Ising_Heis}, we initialize the two-qubit system in the state $\ket{\psi(0)} = (\ket{00} + \ket{01} + \ket{11})/\sqrt{3}$. We choose this initial state to ensure that the two-qubit system evolves through all four components of the computational basis.
We can see that being initialized in $\ket{\psi(0)}$ and with the small leakage, the two-qubit system primarily evolves within the computational basis. The evolution appears to comprise two nearly independent rotations, one involving the components $\ket{00}$ and $\ket{11}$, and the other involving $\ket{01}$ and $\ket{10}$. This behavior arises from the structure of $H_{QQ}$ in \eqref{eq:H_QQ}, which couples $\ket{01}$ and $\ket{10}$ with $\omega_1 = \chi J_{AB}$, and $\ket{00}$ and $\ket{11}$ with $\omega_2 = \sqrt{\omega_1^2 + \Delta^2}$, thus resulting in a faster evolution of the two-qubit system among the latter components.

If we choose two very large background magnetic fields for the two qubits in opposite directions, we can effectively push away all the spin polarised states of the system from the computational basis. In this limit, the coupling effectively becomes an Ising coupling. Therefore, the only product states that can interfere with the computational basis are the pairs made of non-spin-polarised states, where each state belongs to the $S_{\rm tot}^z=0$ subspace of its own chain.
Among these states, the closest to the computational basis are $\ket{SQ_0}$ and $\ket{Q_0S}$, consisting of the nonmagnetic first quintuplet of one chain and the ground state singlet of the other chain. When the two chains are not coupled, these states are separated from the computational basis by $\Gamma-\Delta$.

In this limit, where the coupling is effectively an Ising coupling, one can imagine that the leakage will be smaller for the same coupling energy $J_{AB}$.
To illustrate this point, we also present, with dashed curves, the time evolution of the same two STH qubits now coupled by an Ising interaction instead of the Heisenberg interaction of \eqref{eq:Heis_AB}.
It is evident that with an Ising interaction, the leakage is significantly reduced for the same $J_{AB}$. However, the problem with choosing two very large magnetic fields in the background of the qubits is that it transforms the computational basis into highly excited states. Therefore, such a scenario may not be stable enough for executing a long sequence of unitary operations.
The significantly reduced leakage observed with Ising coupling prompts speculation on effective methods to generate Ising coupling between two quantum dots without relying on strong background magnetic fields. This intriguing question remains open for future investigations.

\section{Conclusion \label{sec:conclusion}}

Starting from a microscopic model, we investigated the coupling of two Singlet-Triplet Haldane( STH ) qubits realized in a quantum dot nanowire with four electrons in each dot through a gated control dot. By computing the energy spectrum of the fermionic system, we demonstrated that each quantum dot nanowire system behaves like a topological Haldane chain with Haldane quasiparticles at both ends. Furthermore, we established the equivalence between controlling the detuning of the control dot and controllable antiferromagnetic coupling between the ends of the two STH qubits. We showed that the low-energy behavior of two coupled STH qubits, effectively made of two spin-half Haldane quasiparticles, resembles that of two simple ST qubits, each made of two real spin-halfs.

Using the spin model, we focused on the issue of decoupling $\ket{T_+T_-}$ states from the computational basis while generating two-qubit gates. We proposed that by using an internal link in each STH qubit, one can vary their length and effectively switch the singlet-triplet gap $\Delta$ on and off, thereby enabling the use of Levy sequence for generating two-qubit gates \cite{levy_two_qb}.

Furthermore, we demonstrated that resolving the leakage problem without altering the length of the STH qubits is attainable by employing different background magnetic fields for the two qubits. Additionally, we showed that inter-chain Ising coupling could substantially decrease leakage, and discussed its potential realization through the use of very large magnetic fields in the background. We anticipate exploring alternative methods for achieving Ising coupling that do not rely on the use of large magnetic fields in the future.
Exploring other avenues for coupling these STH qubits, such as photon bus, remains an interesting direction for future research. We hope this work motivates future experimental work on macroscopic quantum states in semiconductor nanostructures.

\begin{acknowledgments}
We acknowledge NSERC Alliance Quantum Consortium PQS2D grant ALLRP/578466-2022, the QSP-078, AQC-004 and HTSN-341 projects of the Quantum Sensors, Applied Quantum Computing and On-chip integrated circuits based on 2D materials Programs at the National Research Council of Canada, University of Ottawa Research Chair in Quantum Theory of Materials, Nanostructures, and Devices, and Digital Research Alliance Canada with computing resources.
\end{acknowledgments}

\appendix

\section{Two coupled simple ST qubits \label{sec:coupled_ST_appendix}}
Consider $\widehat{H}$ in \eqref{eq:HST_AB} that describes two coupled simple ST qubits. Since $\widehat{H}$ conserves total spin, its spectrum consists of two singlets, three triplets, and one quintuplet. The quintuplet is $\ket{Q_{TT}}\propto 2\ket{T_0T_0}+\ket{T_+T_-}+\ket{T_-T_+}$, and its energy is given by $E_{Q_{TT}}=\Delta+J_{AB}/4$. Next observe that one of the triplets is $\ket{T_{ST+}}\propto \ket{ST_0}+\ket{T_0S}$ and its energy is given by $E_{T_{ST+}}= J_{AB}/4$. This leaves us with two 2D subspaces, one for the singlets and the other for the remaining two triplets. Finding the eigenstates of 2D Hamiltonians is straightforward. We find that the two remaining triplets are $\ket{T_{ST-}} \propto C_1 (\ket{ST_0}-\ket{T_0S}) - J_{AB}(\ket{T_+T_-}-\ket{T_-T_+})$, and $\ket{T_{TT}} \propto J_{AB} (\ket{ST_0}-\ket{T_0S}) + C_1(\ket{T_+T_-}-\ket{T_-T_+})$, with $C_1=\Delta+\sqrt{\Delta^2+J_{AB}^2}$; their energy is given by $E_{T_{ST-}} = -\frac12\sqrt{\Delta^2 + J_{AB}^2} +\Delta/2 - J_{AB}/4$, and $E_{T_{TT}}=\frac12\sqrt{\Delta^2 + J_{AB}^2} +\Delta/2 - J_{AB}/4$. In the limit of $J_{AB}\ll \Delta$, $\ket{T_{ST-}}$ is primarily made of one singlet and one triplet, while $\ket{T_{TT}}$ is primarily made of two triplets.
And finally we find that the two singlets are $\ket{S_{TT}}\propto C_2(\ket{T_+T_-}+\ket{T_-T_+}-\ket{T_0T_0})-3J_{AB}\ket{SS}$, and $\ket{S_{SS}}\propto J_{AB}(\ket{T_+T_-}+\ket{T_-T_+}-\ket{T_0T_0})+C_2\ket{SS}$, with $C_2=2\sqrt{4\Delta^2 - 2\Delta J_{AB} + J_{AB}^2} + 4\Delta-J_{AB}$; their energy is given by $E_{S_{TT}} = \frac12\sqrt{4\Delta^2 - 2\Delta J_{AB} + J_{AB}^2} - J_{AB}/4$, and $E_{S_{SS}}=-\frac12\sqrt{4\Delta^2 - 2\Delta J_{AB} + J_{AB}^2} - J_{AB}/4$. The ground state is $\ket{S_{SS}}$ and in the limit of $J_{AB}\ll \Delta$, it is primarily made of two singlets, while $\ket{S_{TT}}$ is primarily made of two triplets.
This spectrum is shown in Figure~\ref{fig:Heis_N10_coupled} by dashed curves as a function of $J_{AB}/\Delta$.

\section{The structure of coupling matrix \label{sec:matrix_element_appendix}}
Here, we demonstrate that the coupling operator
\begin{align}
\widetilde{H}_{AB} &= \vec{S}_{A,N}\cdot\vec{S}_{B,N} \nonumber \\
&= S_{A,N}^z S_{B,N}^z + \frac{1}{2} (S_{A,N}^+ S_{B,N}^- + S_{A,N}^- S_{B,N}^+),
\label{eq:H_AB}
\end{align}
which appears in \eqref{eq:Heis_AB}, has the same structure as described in Eqs.~(\ref{eq:HQQ_Tpm} and \ref{eq:HST_blocks}) within the subspace spanned by $\lbrace \ket{SS}, \ket{ST_{0}}, \ket{T_{0}S}, \ket{T_{0}T_{0}}, \ket{T_{+}T_{-}}, \ket{T_{-}T_{+}}\rbrace$.

Before starting the proof, observe that we can write a product state of two chains in terms of their configurations as
\begin{equation}
    \ket{ij}=\sum_{p,q}a_{ip}b_{jq}\ket{p;q},
    \label{eq:product_state}
\end{equation}
where we are concerned with $i,j\in \lbrace S,T_0,T_+,T_-\rbrace$, and $p(q)$ are configurations of a single chain of length $N$ like $p = p_1\cdots p_N$, with $p_k\in \lbrace0,\pm 1\rbrace$.
Let us also introduce inverted configuration $\overline{p}$, for which $\overline{p}_k=-p_k$ for all $k$'s.
Then since the Hamiltonian does not distinguish between $\pm \hat{z}$ directions, for all eigenstates of a single chain in its $S_{\rm tot}^z=0$ subspace, we have $|a_{ip}|=|a_{i\overline{p}}|$.

\subsection{The structure of $H_{QQ}$ \label{sec:HQQ}}
For $H_{QQ}$ block we need to show that the only non-zero matrix elements are those located on the anti-diagonal of $H_{QQ}$, corresponding to the terms proportional to $J_{AB}$ in \eqref{eq:H_QQ}.
Notice that in this block the only relevant terms of $\widetilde{H}_{AB}$ is the Ising part $S_{A,N}^z S_{B,N}^z$.
Then using orthonormality of the configurations we can write
\begin{equation}
    \bra{i'j'}S_{A,N}^zS_{B,N}^z\ket{ij}=
    \sum_{p,q}p_Nq_N a_{i'p}^*a_{ip} b_{j'q}^*b_{jq}.
    \label{eq:Ising_matrix_elements}
\end{equation}
If $i=i'$ or $j=j'$ we have $|a_{ip}|^2$ or $|b_{jq}|^2$ in the summand of \eqref{eq:Ising_matrix_elements}.
But since the sum in \eqref{eq:Ising_matrix_elements} goes over all $p$'s and $q$'s, and we have $p_N|a_{ip}|^2 + \overline{p}_N|a_{i\overline{p}}|^2= q_N|b_{jq}|^2 + \overline{q}_N|b_{j\overline{q}}|^2 = 0$, then $\bra{i'j'}S_{A,N}^zS_{B,N}^z\ket{ij}=0$, unless $i\neq i'$ and $j\neq j'$, that is on the anti-diagonal of $H_{QQ}$.
Moreover, since $H_{QQ}$ is Hermitian to complete the proof we just need to show
$\bra{SS}S_{A,N}^zS_{B,N}^z\ket{T_0T_0} = \bra{ST_0}S_{A,N}^zS_{B,N}^z\ket{T_0 S}$.
To see that, we realize that since the Hamiltonian is real, its eigenvectors are also real, therefore
\begin{align}
    \label{eq:chi}
    \bra{SS}S_{A,N}^zS_{B,N}^z\ket{T_0T_0} &= \sum_{p,q}p_Nq_N a_{Sp} a_{T_0 p} b_{S q} b_{T_0 q}\\
    &= \bra{ST_0}S_{A,N}^zS_{B,N}^z\ket{T_0 S}=\chi \nonumber.
\end{align}

\subsection{The structure of $H_{\overline{Q}Q}$ \label{sec:HQq}}
Here, we derive the structure of the matrix elements of $\widetilde{H}_{AB}$ between $\ket{T_+T_-}$ and the computational basis, and the procedure is the same for $\ket{T_-T_+}$. Notice that the only term in $\widetilde{H}_{AB}$, given in \eqref{eq:H_AB}, that produces a non-zero matrix element between $\ket{T_+T_-}$ and the computational basis is $\frac12 S_{A,N}^-S_{B,N}^+$.

Let us start by making a few observations.
First, observe that we have $S_{\rm tot}^\pm\ket{T_0}=\sqrt{2}\ket{T_\pm}$, and $ S_{\rm tot}^\pm\ket{S} = S_{\rm tot}^\pm\ket{T_\pm}=0$, where $S_{\rm tot}^\pm = \sum_n S_{n}^\pm$.
Second, observe that using the commutation relation $[S^+,S^-]=2S^z$, and the first observation, we have $\bra{S}S_N^\mp\ket{T_\pm} = \frac{1}{\sqrt{2}}\bra{S}S_N^\mp S_{\rm tot}^\pm\ket{T_0}=\mp\sqrt{2}\bra{S}S_N^z\ket{T_0}$.
Similarly we also have $\bra{T_0}S_N^\mp\ket{T_\pm} = \frac{1}{\sqrt{2}}\bra{T_\pm}S_{\rm tot}^\pm S_N^\mp \ket{T_\pm}=\pm\sqrt{2}\bra{T_\pm}S_N^z\ket{T_\pm}$.
Here we suppressed the chain index A and B to stay general.

Then, using the second observation, we can write $\bra{SS}\widetilde{H}_{AB}\ket{T_+T_-}=\frac12 \bra{SS}S_{A,N}^-S_{B,N}^+\ket{T_+T_-}=-\bra{SS}S_{A,N}^zS_{B,N}^z\ket{T_0T_0}=-\chi$, which gives us the first column of $H_{\overline{Q}Q}$ in agreement with the form in \eqref{eq:HST_blocks}.

Next, for the next two columns of $H_{\overline{Q}Q}$, it is evident that we should have $|\bra{ST_0}\widetilde{H}_{AB}\ket{T_\pm T_\mp}|=|\bra{T_0 S}\widetilde{H}_{AB}\ket{T_\mp T_\pm}|$, as the Hamiltonian doesn't have preferred direction, and as it is symmetric with respect to the two chains.
To complete the proof, we show that $\bra{ST_0}\widetilde{H}_{AB}\ket{T_+T_-}=\bra{T_0 S}\widetilde{H}_{AB}\ket{T_-T_+}=\chi'$, and the other two matrix elements have the opposite sign.
As mentioned we can write $\ket{T_+T_-}$ as
\begin{subequations}
\begin{align}
    \ket{T_+ T_-} &=\sum_{p,q} a_p b_q \ket{p;q}\\
    \to \ket{T_- T_+} &=\sum_{p,q} a_p b_q \ket{q;p}.
    \label{T+T-_conf}
\end{align}    
\end{subequations}
Consequently, we have
\begin{subequations}
\begin{align}
    \label{eq:SS_T+T-}
    \frac12 S_{A,N}^-S_{B,N}^+\ket{T_+ T_-} & = \sum_{p,q} a_p b_q \ket{p';q'}\\
    \frac12 S_{A,N}^+S_{B,N}^- \ket{T_- T_+} &= \sum_{p,q} a_p b_q \ket{q';p'},
    \label{eq:SS_T-T+}
\end{align}
\end{subequations}
where $p'$ and $q'$ are configurations that we obtain by flipping the last spin down and up in each chain respectively.
It is evident that \eqref{eq:SS_T+T-} and \eqref{eq:SS_T-T+} are the same up to switching between the A and B configurations, which implies that we have $\bra{ST_0}S_{A,N}^-S_{B,N}^+\ket{T_+T_-}=\bra{T_0 S}S_{A,N}^+S_{B,N}^-\ket{T_-T_+}$, as we claimed.

Finally, for the last elements of $H_{\overline{Q}Q}$, using the observations we made above, we can write $\bra{T_0T_0}\widetilde{H}_{AB}\ket{T_+T_-}=\frac12 \bra{T_0T_0}S_{A,N}^-S_{B,N}^+\ket{T_+T_-}=-\bra{T_+T_-}S_{A,N}^zS_{B,N}^z\ket{T_+T_-}=\chi''$, which proves full agreement with \eqref{eq:HST_blocks}.

\subsection{The structure of $H_{\overline{QQ}}$ \label{sec:Hqq}}
The structure of $H_{\overline{QQ}}$ is simple.
It is clear that it should be proportional to the identity matrix, as $\widetilde{H}_{AB}$ cannot connect $\ket{T_+ T_-}$ and $\ket{T_- T_+}$, because such a process would require two spin flips in each chain. Moreover, since the Hamiltonian does not distinguish between the $\pm \hat{z}$ directions, we should have $\bra{T_+ T_-} \widetilde{H}_{AB} \ket{T_+ T_-} = \bra{T_- T_+} \widetilde{H}_{AB} \ket{T_- T_+}= -\chi''$, in agreement with \eqref{eq:HST_blocks}.

\section{Generating CNOT gate \label{sec:CNOT_appendix}}
Here we show how to generate a gate {\em locally equivalent} to CNOT using $H_{QQ}$ in \eqref{eq:H_QQ}.
We say two two-qubit gates $U$ and $U'$ are locally equivalent if $U' = (L_A\otimes L_B) U (R_A\otimes R_B)$, where $L$'s and $R$'s are single-qubit gates acting on qubits A and B. It can be shown that if two two-qubit gates have the same {\em canonical form} $\alpha=(\alpha_1,\alpha_2,\alpha_3)$ in the {\em Weyl chamber} $\frac{\pi}{4} \geq \alpha_1 \geq \alpha_2 \geq |\alpha_3| \geq 0 $, then they are locally equivalent \cite{canonical_form}.
There is a standard procedure for finding the canonical form of a two-qubit gate like $U$, as the following \cite{canonical_form}: (1)~Factor out a global phase $U = e^{i\alpha_0}\widetilde{U}$ such that $\det(\widetilde{U})=1$. (2)~Form $V=M^\dagger \widetilde{U} M$ where
\begin{equation}
	M = \frac{1}{\sqrt{2}} \begin{bmatrix}
	1 & 0 & 0 & i \\
	0 & i & 1 & 0 \\
	0 & i & -1 & 0 \\
	1 & 0 & 0 & -i
	\end{bmatrix}.
	\label{eq:magic}
\end{equation}
(3)~Find $\lambda$'s, the eigenvalues of $V^TV$. (4)~Then the canonical form $\alpha$ is the unique solution of the following set of equations in the Weyl chamber
\begin{subequations}
	\begin{align}
		\lambda_1 &= e^{2i(\alpha_1 - \alpha_2 + \alpha_3)},\\
		\lambda_2 &= e^{2i(\alpha_1 + \alpha_2 - \alpha_3)},\\
		\lambda_3 &= e^{-2i(\alpha_1 + \alpha_2 + \alpha_3)},\\
		\lambda_4 &= e^{-2i(\alpha_1 - \alpha_2 - \alpha_3)}.
	\end{align}
	\label{eq:eigen_alpha}
\end{subequations}

The canonical form of CNOT -- along with many other common two-qubit gates -- is $\alpha_{\rm CNOT} =  (\frac\pi4,0,0)$.
Next, we show how to use $H_{QQ}$ in \eqref{eq:H_QQ} to generate a gate with the same canonical form.

\subsection{Arbitrary coupling value \label{sec:any_J_AB}}
Here we show how to find $t_c$ such that for an arbitrary coupling $J_{AB}$, $U_{2QQ}(t_c)=e^{-it_c H_{QQ}} (XI) e^{-it_c H_{QQ}}$ has a canonical form the same as $\alpha_{\rm CNOT}$.
Using the procedure mentioned above, we observe that for any $t$, the canonical form of $U_{2QQ}(t)$ has the form $\alpha_{2QQ}(t)=(\phi,0,0)$, where
\begin{subequations}
\begin{align}
	\sin(\phi) &= \text{abs}\big( \sin(\omega_1 t)\cos(\omega_2 t) \nonumber\\
    & + \frac{\sin(\omega_2 t)}{\omega_2}
	(\omega_1 \cos(\omega_1 t) + i \Delta \sin(\omega_1 t))\big),\\
	\cos(\phi) &=  \text{abs}\big(\cos(\omega_1 t)\cos(\omega_2 t) \nonumber\\
    & +i \frac{\sin(\omega_2 t)}{\omega_2}
	(\Delta\cos(\omega_1 t) + i \omega_1 \sin(\omega_1 t))\big),
\end{align}
\label{eq:gamma}
\end{subequations}
with $\omega_1 = \chi J_{AB}$, and $\omega_2 = \sqrt{\omega_1^2 + \Delta^2}$.
Therefore, to find $t_c$ one needs to solve $\sin^2(\phi(t_c)) = \cos^2(\phi(t_c)) = \frac12$, which can be done numerically, for any given $J_{AB}$.

\subsection{Special coupling values \label{sec:special_J_AB}}
It turns out that for a set of special coupling values $J_{AB}$, one can generate a gate locally equivalent to CNOT by one time application of $H_{QQ}$ in \eqref{eq:H_QQ}, as $U_{QQ}(t_c)=e^{-it_c H_{QQ}}$.
To see that, we observe that the canonical form of $U_{QQ}(t)$ has the form $\alpha_{QQ}(t) = \frac{1}{2}(\phi_1+\phi_2, \phi_1-\phi_2, 0)$, where $\phi_1=\omega_1 t$, and $\sin(\phi_2) = \frac{\omega_1}{\omega_2}\sin(\omega_2 t)$.
Therefore, $\alpha_{QQ}$ can reach at the $\alpha_{\rm CNOT}$ point, if we have $\phi_1 + \phi_2 = \frac\pi2 + m_1\pi$, and $\phi_1 - \phi_2=m_2\pi$, with integer $m$'s. This implies that $\phi_1 = (2m + 1)\frac\pi4$, and that $\sin^2(\phi_2)=\frac12$.
Finding $t$ from $\phi_1$ and plugging it in the expression for $\sin(\phi_2)$ we find
\begin{equation}
    2\sin^2\left((2m+1)\frac{
    \pi}{4}\frac{\omega_2}{\omega_1}\right) = \left(\frac{\omega_2}{\omega_1}\right)^2.
    \label{eq:special_J_equation}
\end{equation}
For integers $m>1$, this equation has a set of solutions for $\omega_2/\omega_1$, from which the special values of $J_{AB}$ for a given $\Delta$ can be obtained.

\bibliography{references}

\end{document}